\documentclass{elsart}

\usepackage{natbib}

\newcommand{\arcmins}{\mbox{$^{\prime}$}}
\newcommand{\parcsec}{\mbox{$\stackrel{\prime\prime}{\textstyle .}$}}

\newcommand{\psec}{\mbox{$\stackrel{s}{\textstyle .}$}}
\newcommand{\hours}{\mbox{$^{h}$}}
\newcommand{\mins}{\mbox{$^{m}$}}

\usepackage{graphicx}

\journal{New Astronomy}

\usepackage{amsmath,amssymb}

\begin{document}

\begin{frontmatter}


\vspace*{-1.1cm}
\title{Small-scale variations in the radiating surface of the 
 GRB~011211 jet}
\author[label1,label2]{P.~Jakobsson\corauthref{cor1}},
\ead{pallja@astro.ku.dk}
\author[label1]{J.~Hjorth},
\author[label3]{E.~Ramirez-Ruiz},
\author[label4]{C.~Kouveliotou},
\author[label1]{K.~Pedersen},
\author[label1,label5]{J.~P.~U.~Fynbo},
\author[label6,label22]{J.~Gorosabel},
\author[label1]{D.~Watson},
\author[label1]{B.~L.~Jensen},
\author[label7]{T.~Grav},
\author[label7]{M.~W.~Hansen},
\author[label1]{R.~Michelsen},
\author[label8]{M.~I.~Andersen},
\author[label5,label33]{M.~Weidinger},
\author[label1]{H.~Pedersen}

\corauth[cor1]{Corresponding author.}

\address[label1]{Niels Bohr Institute, Astronomical Observatory, 
  University of Copenhagen, Juliane Maries Vej 30, 
  DK-2100 Copenhagen, Denmark}
\address[label2]{Science Institute, University of Iceland, Dunhaga 3,
  107 Reykjav\'{\i}k, Iceland}
\address[label3]{Institute of Astronomy, University of Cambridge, 
  Madingley Road, Cambridge, CB3 0HA, England, UK}
\address[label4]{NSSTC, SD-50, 320 Sparkman Drive, Huntsville, Alabama
  35805, USA}
\address[label5]{Department of Physics and Astronomy, University of
  Aarhus, Ny Munkegade, 8000 \AA rhus C, Denmark}
\address[label6]{STScI, 3700 San Martin
  Drive, Baltimore, MD 21218, USA}
\address[label22]{Instituto de Astrof\'{\i}sica de Andaluc\'{\i}a 
(IAA-CSIC), P.O. Box 03004, E-18080 Granada, Spain}
\address[label7]{Institute of Theoretical Astrophysics, University of
  Oslo, PB 1029 Blindern, 05315 Oslo, Norway}
\address[label8]{Astrophysikalisches Institut Potsdam, An der
  Sternwarte 16, 14482 Potsdam, Germany}
\address[label33]{European Southern Observatory, 
  Karl-Schwarzschild-Stra\ss e 2, 85748, Garching bei M\"unchen, Germany}
\begin{abstract}
We report the discovery of the afterglow of the X-ray
rich, long-duration gamma-ray burst GRB~011211 and present evidence
for oscillatory behaviour in its early optical light curve. 
The time-scale of the fluctuations, $\sim$1 hour, is much 
smaller than the time of the observations, $\sim$12 hours from 
the onset of the gamma-ray burst. The character and strength of 
the fluctuations are unprecedented and are inconsistent with causally 
connected variations in the emission of a symmetric, relativistic 
blast wave, i.e. flux variations which are produced 
uniformly throughout the shell surface are ruled out. Therefore, 
the wiggles are the result of spherically asymmetric density or energy
variations. Additionally, there is evidence for fluctuations in the 
X-ray afterglow light curve. If real, the resulting difference in the
observed time of the peaks of the short-term
variations at X-ray and optical frequencies, would demonstrate that 
the energy content across the jet-emitting surface is not uniform.
\end{abstract}


\end{frontmatter}


\section{Introduction}
GRB 011211 was detected on 2001 December 11.798 UT \citep{gand} 
with one of the Wide Field Cameras (WFCs) on board the 
Italian-Dutch satellite Beppo\-SAX. The prompt gamma-ray 
emission lasted $\sim$270~s, making it one of the 
longest bursts observed with the satellite. Following the 
distribution of a 2\arcmins\ radius error circle \citep{gand2}, we 
identified \citep{grav} the optical counterpart of the gamma-ray burst
(GRB). An X-ray afterglow was subsequently detected \citep{santos} 
and its redshift measured via absorption lines in the optical spectrum 
to be $z=2.140$ \citep{andy,holland}. The analysis of 
the X-ray spectrum \citep{james} showed emission
lines arising in metal-enriched material with an outflow velocity of
$\sim$0.1$c$. The presence of such line features strongly
suggests a massive stellar progenitor, but the details remain model
dependent \citep{lazzati}.
\par
Afterglow observations of long-duration GRBs typically show 
their flux to decline as a power-law in time. However, recent
continuous early-time monitoring of GRB optical light curves has shown  
evidence for variations or bumps superposed on power-law decays
\citep{bersier,fox,japan}. We note that these are not supernova 
bumps as observed in the optical afterglow (OA) light curves of
several bursts \citep[see e.g.][]{bloom,garnavich,jens2}. We present 
here for the first time a comparative study of multi-wavelength 
oscillations observed in the light curve of GRB~011211. 
Despite intensive monitoring of many GRB counterparts 
\citep{jens,stanek,halpern,burenin,ls,jav}, the only other burst for 
which short-term time-scale (less than one hour) 
variations have been detected in its optical light curve is 
GRB 021004, albeit in a random fashion \citep{bersier}. 
The oscillatory behaviour and the rapid, sharp decline and rise of 
the pulses of the light curve of GRB~011211 presented here, are unlike 
any other light curve reported before. Our observations provide a useful 
probe of the GRB explosion, the structure of the 
emerging jet, as well as the structure of the surrounding environment. 
\begin{figure}
   \centering
   \includegraphics[width=0.48\textwidth]{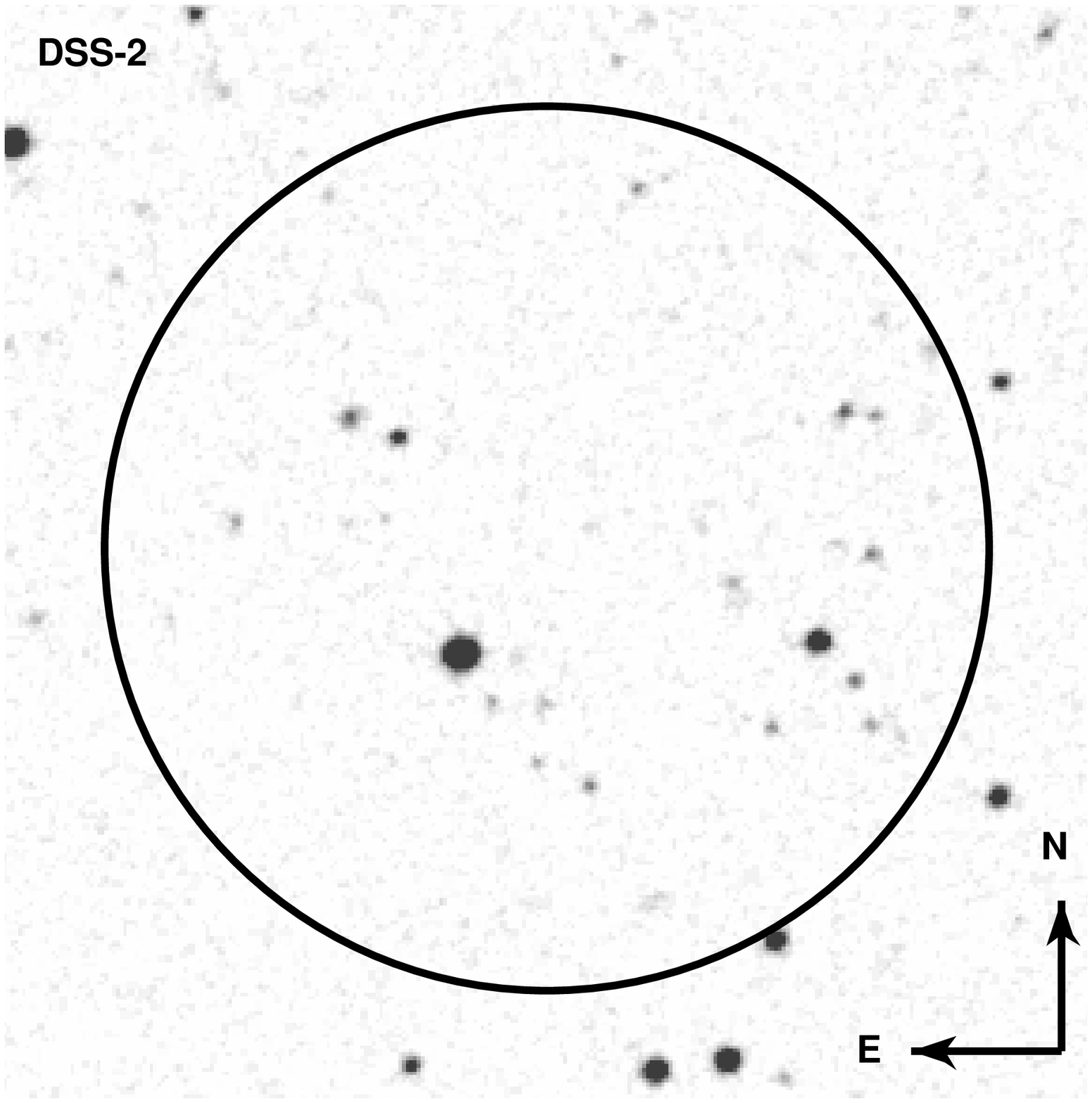}
   \includegraphics[width=0.50\textwidth]{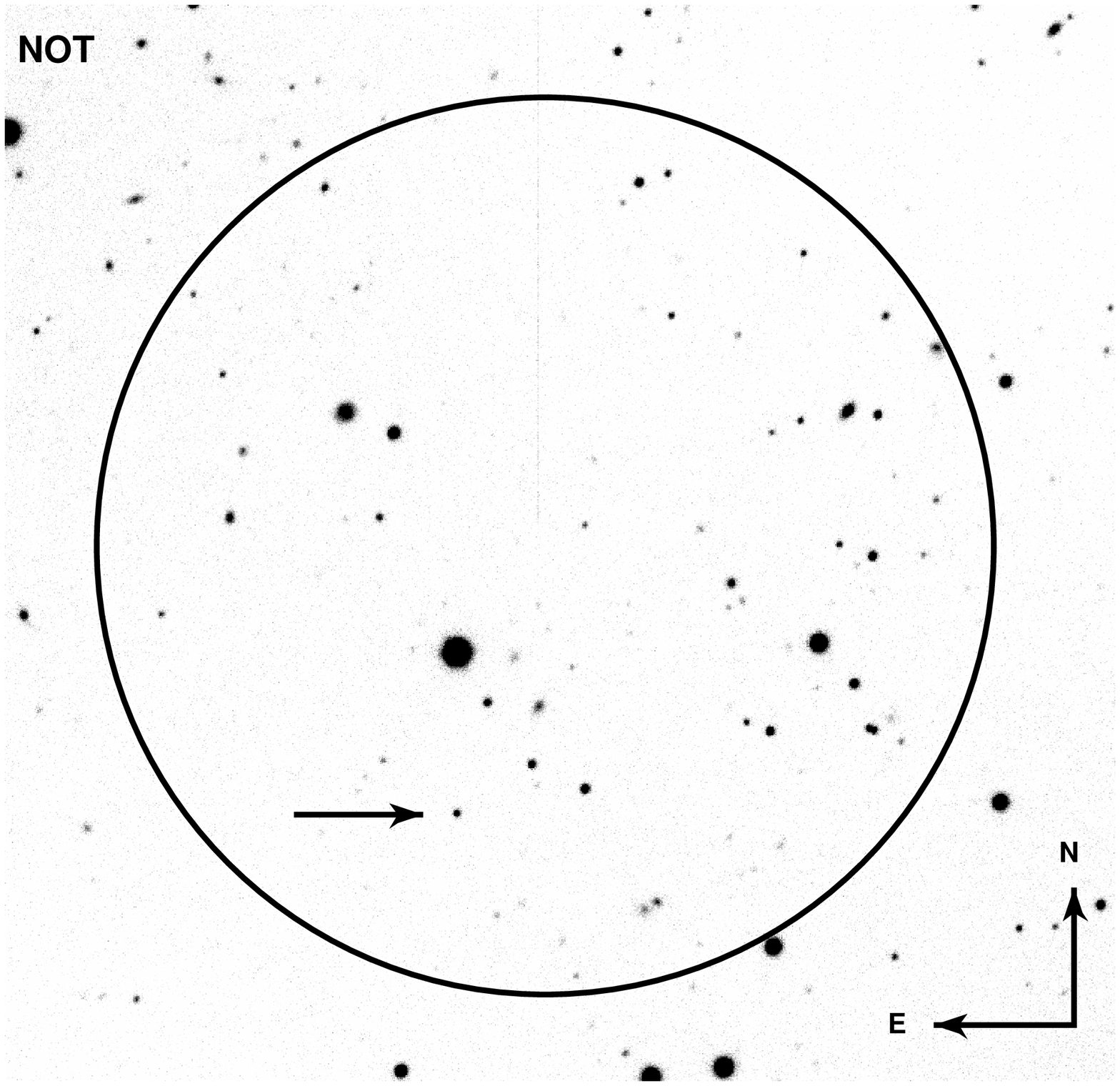}
\caption{\small \em Images of the field of the optical transient of 
GRB~011211. \emph{Left:} The $R$-band DSS-2 image. 
\emph{Right:} The discovery $R$-band 
image of the GRB~011211 optical afterglow (OA), taken with 
the NOT. The position of the OA is marked with an arrow. 
The BeppoSAX localization \citep{gand2} is indicated by a 
circle of radius $2\arcmins$. Both images cover a square of 
$5\arcmins \times 5\arcmins$.}
   \label{uno.fig}
\end{figure}
\section{Observations \& data analysis}
We observed GRB 011211 $\sim$9.6 hours after 
the GRB trigger (starting 2001 December 12.22 UT) with the 
Nordic Optical Telescope (NOT) at the Canary Island of La Palma 
and with the Danish 1.54-m telescope at La Silla in Chile, in 
standard $B$-, $V$-, $R$-, and $I$-bands \citep{grav}. 
A detailed log of our observations is listed in 
\mbox{Table~\ref{obs.tab}}. A comparison of our first 
image with the Digitized Sky Survey 2 (DSS-2) frames, identified 
a new source shown in Fig.~\ref{uno.fig}. Subsequent observations with the 
NOT revealed that the source was fading, thus establishing that the 
nature of the transient was consistent with being the OA of GRB 011211. 
\par
Reference stars for astrometry were collected using the 
large field of view provided by the Danish Faint Object Spectrograph 
and Camera (DFOSC) mounted on the Danish 1.54-m 
($13.7 \times 13.7$~arcmin$^2$). Based on $\sim$50 
USNO stars per image, we 
determined 11 independent positions of the afterglow. The mean 
value of the 11 afterglow coordinates is RA(J2000) = 
$11\hours15\mins17\psec98$ and 
Dec(J2000) = $-21^{\circ}56\arcmins56\parcsec1$ 
with an error of $0\parcsec3$.
\par
\begin{table}
\caption{\small \em Photometry of the afterglow of GRB~011211. 
The magnitude 
of the optical afterglow was measured relative to eight stars in 
the field. The calibrated magnitudes of these stars are given in 
J03. Due to saturation we only used a subset of these stars in some 
of our images. The magnitudes were calculated using point-spread 
function fitting photometry.}
  \centering
  \setlength{\arrayrulewidth}{0.8pt}   
  \begin{tabular}{ccccc}
  \vspace{-4 mm} \\
  \hline
  \hline
Date (UT)  & Tel. & Magnitude & Seeing   & Exp. time \\
(2001 Dec) &      &           & (arcsec) & (s) \\
\hline
\hspace{-5 mm}
\emph{B-band:}    &     &                  &      & \\
12.2816  & NOT    & $ 21.229 \pm 0.047$ & 1.1 &   300 \\
12.3539  & 1.54-m & $ 21.262 \pm 0.089$ & 1.4 &   200 \\
\hspace{-5 mm}
\emph{V-band:}    &     &                  &      & \\
12.2907  & NOT    & $ 20.761 \pm 0.080$ & 1.0 &   300 \\
12.3586  & 1.54-m & $ 20.987 \pm 0.042$ & 1.2 &   300 \\
\hspace{-5 mm}
\emph{R-band:}    &     &                  &      & \\
12.2181  & NOT    & $ 20.012 \pm 0.033$ & 1.7 &   300 \\
12.2227  & NOT    & $ 19.877 \pm 0.029$ & 1.4 &   300 \\
12.2272  & NOT    & $ 19.918 \pm 0.022$ & 1.3 &   300 \\
12.2831  & 1.54-m & $ 20.368 \pm 0.040$ & 1.5 &   600 \\
12.2912  & 1.54-m & $ 20.298 \pm 0.033$ & 1.5 &   600 \\
12.2994  & 1.54-m & $ 20.230 \pm 0.038$ & 1.4 &   600 \\
12.3075  & 1.54-m & $ 20.285 \pm 0.042$ & 1.4 &   600 \\
12.3157  & 1.54-m & $ 20.274 \pm 0.051$ & 1.3 &   600 \\
12.3238  & 1.54-m & $ 20.330 \pm 0.046$ & 1.1 &   600 \\
12.3633  & 1.54-m & $ 20.539 \pm 0.063$ & 1.1 &   300 \\
12.3681  & 1.54-m & $ 20.616 \pm 0.100$ & 1.0 &   300 \\
\hspace{-5 mm}
\emph{I-band:}    &     &                  &      & \\
12.2861  & NOT    & $ 19.965 \pm 0.110$ & 0.9 &   300 \\
12.3350  & 1.54-m & $ 19.951 \pm 0.054$ & 1.3 &   300 \\
12.3397  & 1.54-m & $ 19.981 \pm 0.071$ & 1.8 &   300 \\
12.3491  & 1.54-m & $ 20.064 \pm 0.061$ & 1.4 &   300 \\
\hline
\end{tabular}
  \label{obs.tab}
\end{table}
\begin{figure}
   \centering
   \includegraphics[width=0.99\textwidth]{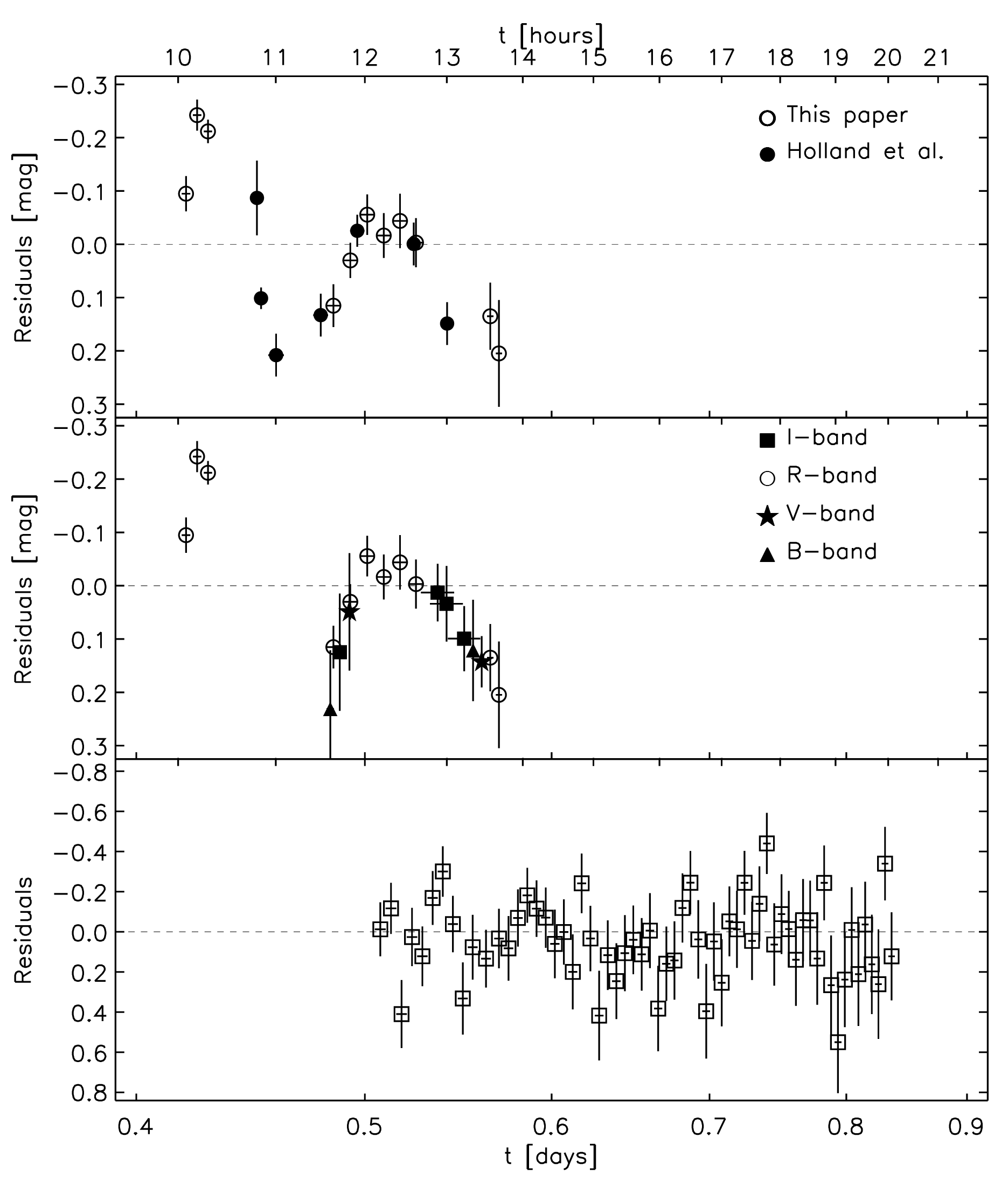}
\caption{\small \em Comparison of the residuals in the optical and 
X-ray light curves of the afterglow of GRB 011211. \emph{Top:} The 
$R$-band data are drawn from Table~\ref{obs.tab} and the 
literature \citep{holland}. \emph{Middle:} 
The same as in the top panel, except here we also plot the 
$I$-, $V$-, and $B$-band residuals 
(data taken from Table~\ref{obs.tab}). Offsets relative to 
the $R$-band were applied using color 
information from the spectral energy distribution (J03). 
\emph{Bottom:} The XMM-Newton X-ray light curve residuals 
(0.5--10~keV). In order to compare the X-ray results with 
the optical, the units are expressed in 
terms of $-2.5 \times \log(\textrm{flux})$. In all three 
panels, the horizontal error bars are equal to the integration 
time for each data point, and $t$ is the time since 
the burst.}
   \label{dos.fig}
\end{figure}
We obtained very well sampled early $R$-band observations of the OA 
$\sim$10--14 hours after the burst. Our data indicated significant 
temporal variations superposed on the power-law fit to the 
light curve of the GRB \citep[][hereafter J03]{palli}. To 
enhance the visibility of these features, we subtracted from our data 
the power-law decay contribution ($\alpha_1 = -0.95$) in 
the optical flux; the result is shown in Fig.~\ref{dos.fig} 
(top panel). The deviation of the data points from a pure 
power-law decline 
is significant ($>$$5\sigma$) as derived from Spearman's rank 
correlation. The residual weighted mean is $0.01 \pm 0.03$ with a 
root mean square (rms) of $13\%$ (after correcting for the 
photon noise). The amplitude of the largest oscillation 
is $\sim$0.25 magnitudes. We detect two maxima with a single 
minimum in-between and by fitting Gaussians to the data set 
we measure the time-scale of the temporal oscillations to 
be $\Delta t_{R} = 0.9 \pm 0.2$ hours. In the middle panel 
of Fig.~\ref{dos.fig} we display the combined multicolor light 
curve of GRB~011211; a wave-like behaviour similar to that in the 
$R$-band light curve is evident in the combined $I$-, $V$-, and $B$-band 
residuals.
\par
The bottom panel of Fig.~\ref{dos.fig} 
shows the temporal variations present in the residuals of the 
X-ray afterglow of GRB~011211 observed with the X-ray Multi 
Mirror-Newton satellite (XMM-Newton) \citep{santos}. The reduction 
of the X-ray data set is described in detail in J03. We observe 
significant deviations from the power-law decline 
($\alpha_{\mathrm{X}} = -1.62$) in the X-ray 
flux ($>$3$\sigma$ as derived from Spearman's rank
correlation). Although 
the X-ray fluctuations are not as significant as the optical ones, 
the value of the (X-ray) Spearman rank correlation coefficient 
translates to a probability of less than $3\times10^{-3}$ that this 
correlation is a statistical fluctuation. 
The weighted mean of the residuals is $0.00 \pm 0.19$, with an 
$\sim$11$\%$ rms variation around the mean flux (after correcting 
for the photon noise). There is weak evidence that the fluctuations
are of a similar oscillatory nature as the optical ones.
The fluctuation time-scale is comparable to 
that observed in the 
optical, with $\Delta t_{\textrm{X}} / \Delta t_R = 0.75 \pm 0.20$.  
From the short overlap of the optical 
and X-ray data, we find that the oscillations are not in 
phase with a time difference of at least $0.70 \pm 0.15$ 
hours (as derived from the cross correlation function). 
\par
\section{Discussion}
The short-term wave-like behaviour of the GRB 011211 optical 
light curve reported here is unprecedented.
This behaviour was not detected in the earlier analysis
of the OA of GRB~011211 reported by \citet{holland},
due to insufficient data sampling. Moreover, the variations detected
in the optical afterglow of GRB~021004 \citep{bersier} appeared 
highly random and erratic, and all other GRBs have displayed 
variations with much longer time-scales. We discuss below the 
origin of the bumps in the optical light curve of GRB~011211.
\par 
The simplest afterglow model, where a
relativistic jet decelerates as it expands into the ambient matter
leading to a radiative output with a characteristic power-law decay,
has been remarkably successful so far in accommodating the present
data \citep{pkw}. How does this model account for the unusual 
optical light
curve of GRB 011211? The strong temporal variations in the early
afterglow could be interpreted as a result of \emph{i)} refreshed
shocks created as the leading edge of the jet decelerates and is
caught up by slower-moving jet gas \citep[e.g.][]{rm98,pmr98,rmr01},
\emph{ii)} the relativistic jet impacting an external medium of
variable density \citep[e.g.][]{wl,enrico,lazzati2,heyl,nakar}, or
\emph{iii)} a non-uniform jet structure
\citep[e.g.][]{mes98,kumar}. We note that a smooth jet traveling
through a clumpy medium would quickly cease to be homogeneous, i.e.
it is possible that \emph{ii)} could give rise to \emph{iii)}.  
Refreshed shocks covering a large fraction of the emitting surface can 
only increase the energy of the blast wave, and therefore cannot account 
for the rapid decay seen in the optical data
(top panel of Fig.~\ref{dos.fig}). 
\par 
The sharpness of the features observed at optical frequencies is
difficult to reconcile with a density discontinuity
covering most of the visible surface of the jet \citep{np}. On the 
other hand, the short time-scale of the oscillations provides 
interesting upper limits of $\sim$1--10~AU on the size of the 
clumps around the source. These limits are
lower than the fluctuation amplitudes seen on similar scales in the
local interstellar medium \citep{wl}, though they may reflect the
length scale of comet-like clumps observed in ring nebulae surrounding
massive stars \citep[e.g.][]{gg}.
\par
\begin{figure}[!h]
   \centering
   \includegraphics[width=0.45\textwidth]{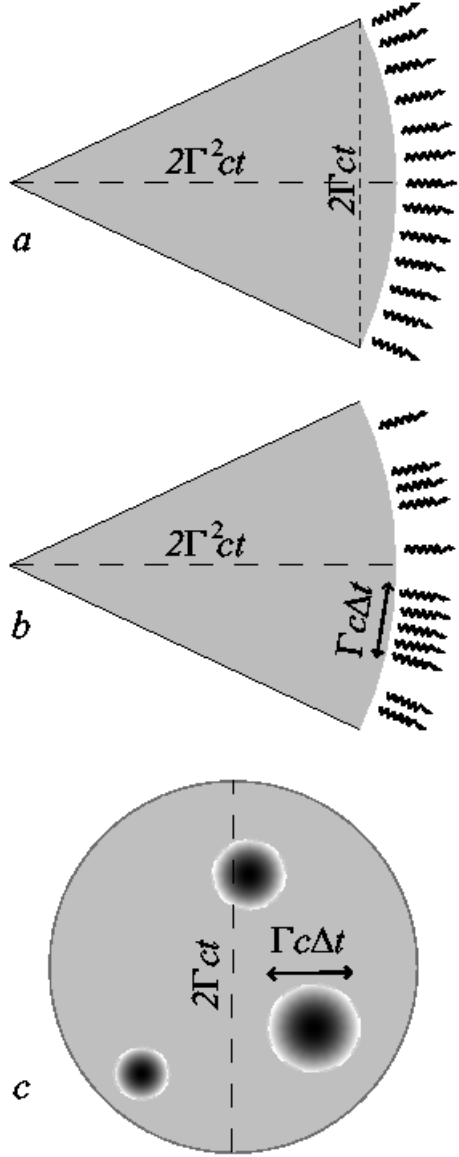}
\caption{\small \em GRB tomography. The emission from a radiating 
object moving close to the speed of light is beamed into a very 
narrow cone. \emph{(a)} A relativistic jet 
that radiates uniformly would have an emitting region
with a transverse size of $2 \Gamma c t$. The afterglow profile would
be the sum of many emitting regions, and thus, should be 
smooth. \emph{(b)} Only regions the size of $\Gamma c \Delta t$ can
produce afterglow fluctuations with a time structure of $\Delta
t$. The same jet pattern observed face-on is shown in 
\emph{(c)}. The photon flux emitted per 
unit solid angle and per unit time at this flare site 
should be larger than that arising from the rest of
the blast wave. The corresponding emission would then average over the
observed region and it may increase (or decrease) depending on the
intrinsic properties of different flare sites across the jet
surface. For example, a single region with a $\Gamma$ that is ten
times larger than for other regions could produce emission that doubles
the overall count rate in a different energy band.}
   \label{tres.fig}
\end{figure}

 
A promising alternative
interpretation is energy variations within the expanding jet. It is
possible for a large fraction of the emitting surface of the jet to
become active and the flare region to remain small, because the energy
content of the jet varies strongly as a function of angle. At the time
$t$ of the observations the region contributing to the total observed flux
would have a transverse size of $2 \Gamma c t$, where $\Gamma$ is the
Lorentz factor of the jet. Only regions the size of $\Gamma c \Delta
t$ can produce afterglow fluctuations with time structures of $\Delta
t \ll t$. In other words, the jet-emitting surface is peppered with many
regions (hot spots) on a scale much smaller than the narrow cone
visible along the line of motion (see Fig.~\ref{tres.fig}). Variations in 
the initial conditions as a function of the opening angle could spread the 
causally disconnected regions out along the line of sight to the
observer such that the emission would arrive at different times.
The corresponding overall emission at a given wavelength is then 
averaged over the observed
region, and it may increase or decrease locally depending on the
number, intensity, and spectrum of the different flare sites across
the jet surface. This can happen either because the radiating material
is in narrow fingers or, as the jet slows down, we observe afterglow
emission not just from material moving directly towards us, but also from a
wider range of angles.  Thus, an inhomogeneous energy jet is a viable
explanation for the observed rapid variations in the light curve, with the
size of a hot spot determining the duration of the bump.
\par 
From our optical observations alone we can safely conclude that the wiggles
in the light curve of GRB~011211 are the result of \emph{spherically
asymmetric density or energy variations}, i.e. variations that cover
less than the observed $1/ \Gamma$ region.\footnote[1]{This is because the
observed time-scale of the variations ($\sim$1~hour) is much shorter
than the overall elapsed time after the burst trigger
($\sim$12~hours), when angular smoothing would have smeared out the
light curve variations.} The question remains: Can we decide between 
the above two scenarios. In principle, simultaneous multi-wavelength
observations can help settle this question (see Salmonson \& Ramirez-Ruiz, in
preparation, for a detailed treatment) and, although the X-ray variations
are of lower significance than the
optical, in what follows we explore the implications of fluctuations
in both bands. As shown by J03,
the X-ray band is located above the cooling frequency at the time of our
observations, where density variations of the ambient medium have a much
weaker impact on the emitted flux 
\citep[see e.g. equation~7 in][]{nakar}.
Given the observed $\sim$11$\%$ variation in the X-ray flux, 
we would then expect a much stronger {\it correlated} optical one, 
which we do not observe. On the other hand, uncorrelated fluctuations 
at different energy bands can be produced by hot
spots (small regions with higher energy and/or $\Gamma$). Since 
$\Gamma$ strongly affects the synchrotron peak frequency, a
situation can easily arise where we observe a hot spot releasing 
most of its energy in the X-rays, without a significant optical 
contribution above the continuum. Conversely, an optical/infrared 
bump can be detected without an X-ray counterpart, at least when the 
cooling frequency of the afterglow continuum spectrum is located between
the optical and the X rays. Thus, the combined departure from a 
power-law decay observed in the optical and X-ray light curves 
seems to disfavour a scenario in which the density of the ambient 
medium varies and the energy is kept constant across the jet surface.
\par 
The above reasoning assumes that the X-ray and optical photons are 
emitted from the surface of the expanding jet, an assumption 
justified by the fact that the variations seen at different energies 
have similar time-scales. Interestingly, the analysis of the X-ray 
spectrum during this time frame uncovered the presence of emission 
line features arising from metal-enriched material \citep{james}. 
It is feasible that a fraction of X-ray photons reach the observer 
after being reflected by material away from the line of sight
\citep{lazzati}. While the presence of such spectral features is not
conclusive \citep[but see e.g.][]{darach}, the X-ray wiggles observed
here could be attributed to variations in the line emission intensity,
which in turn are caused by changes in the illuminating continuum.  The
X-ray afterglow could then, at least in part, be due to the continuing
power output from the GRB central engine interacting with the envelope
of a massive progenitor star \citep{rees}.  
\par Collapsar \citep{mwh}
or magnetar-like \citep{wheeler} models not only provide a natural
scenario for a sudden burst followed by a decaying energy output but
could also easily imprint a non-uniform structure to the GRB jet as it
bores its way through the stellar mantle \citep{zhang}. Further data
on early X-ray spectral and temporal features provided by
the Swift satellite will offer additional clues to the nature of the
progenitor and the relativistic jet that it triggers.

\section*{Acknowledgements}
It is a pleasure to thank Sir M. Rees, S. E. Woosley, G.  Bj\"ornsson,
E. H. Gudmundsson, A. Loeb, S. T. Holland and E. Nakar for 
helpful comments and suggestions. We are grateful to the anonymous 
referee for stimulating remarks which have improved the paper. 
Our results are based on
observations made with the Nordic Optical Telescope, operated on the
island of La Palma jointly by Denmark, Finland, Iceland, Norway, and
Sweden, in the Spanish Observatorio del Roque de los Muchachos of the
Instituto de Astrofisica de Canarias. We acknowledge the availability
of DSS-2 exposures used in this work. P.J. gratefully acknowledges
support from The Icelandic Research Fund for Graduate Students, and a
Special Grant from the Icelandic Research Council. E.R. thanks
CONACyT and the ORS for sponsorship. J.P.U.F. and K.P. acknowledge
support from the Carlsberg foundation. This
work was supported by the Danish Natural Science Research Council
(SNF).  The authors acknowledge benefits from collaboration within the
EU FP5 Research Training Network ``Gamma-Ray Bursts: An Enigma and a
Tool''.



\begin{thebibliography}{}


\bibitem[Bersier et al.(2003)]{bersier} Bersier, D., et al., 2003. 
ApJ 584, L43

\bibitem[Bloom et al.(1999)]{bloom} Bloom, J. S., et al., 1999. 
Nature 401, 453

\bibitem[Burenin et al.(2003)]{burenin} Burenin, R., et al., 2003. 
AstL 29, 573

\bibitem[Fox et al.(2003)]{fox} Fox, D. W., et al., 2003. Nature 422, 284

\bibitem[Fruchter et al.(2001)]{andy} Fruchter, A., Vreeswijk, P.,
  Rhoads, J., Burud, I., 2001. GCN Circ. 1200

\bibitem[Gandolfi(2001a)]{gand} Gandolfi, G., 2001a. GCN Circ. 1188

\bibitem[Gandolfi(2001b)]{gand2} Gandolfi, G., 2001b. GCN Circ. 1189

\bibitem[Garcia-Segura et al.(1996)]{gg} Garcia-Segura G.,
  Langer N., Mac Low M.-M., 1996. A\&A 316, 133

\bibitem[Garnavich et al.(2003)]{garnavich} Garnavich, P., et al., 
2003. ApJ 582, 924

\bibitem[Gorosabel et al.(2003)]{jav} Gorosabel, J., et al., 2003. 
A\&A, submitted (astro-ph/0309748)

\bibitem[Grav et al.(2001)]{grav} Grav, T., et al., 2001. GCN Circ. 1191

\bibitem[Halpern et al.(2003)]{halpern} Halpern, J. P., Mirabal, N.,
  Bureau, M., Fathi, K., 2003. GCN Circ. 2021 

\bibitem[Heyl \& Perna(2003)]{heyl} Heyl, J. S., Perna, R., 2003. 
ApJ 586, L13 

\bibitem[Hjorth et al.(1999)]{jens} Hjorth, J., et al., 1999. 
GCN Circ. 320

\bibitem[Hjorth et al.(2003)]{jens2} Hjorth, J., et al., 2003. 
Nature 423, 847

\bibitem[Holland et al.(2002)]{holland} Holland, S., et al., 2002. 
AJ 124, 639

\bibitem[Jakobsson et al.(2003)]{palli} Jakobsson, P., et al., 2003. 
A\&A 408, 941 (J03)

\bibitem[Kumar \& Piran(2000)]{kumar} Kumar, P., Piran, T., 2000. 
ApJ 535, 152

\bibitem[Laursen \& Stanek(2003)]{ls} Laursen, L. T., Stanek, K. Z., 
2003. ApJ 597, L107

\bibitem[Lazzati et al.(2002a)]{lazzati} Lazzati, D., Ramirez-Ruiz, E., 
Rees, M.~J., 2002a. ApJ 572, L57

\bibitem[Lazzati et al.(2002b)]{lazzati2} Lazzati, D., Rossi, E., 
  Covino, S., Ghisellini, G., Malesani, D., 2002b. A\&A 396, L5  

\bibitem[MacFadyen et al.(2001)]{mwh} MacFadyen,~A.~I., 
Woosley,~S.~E., Heger,~A., 2001. ApJ 550, 410

\bibitem[M\'esz\'aros et al.(1998)]{mes98}
M\'esz\'aros, P., Rees, M. J., Wijers, R., 1998. ApJ 499, 301

\bibitem[Nakar \& Piran(2003)]{np} Nakar, E., Piran, T., 2003. 
ApJ 598, 400

\bibitem[Nakar et al.(2003)]{nakar} Nakar, E., Piran, T., 
Granot, J., 2003. NewA 8, 495

\bibitem[Panaitescu et al.(1998)]{pmr98}
Panaitescu, A., M\'esz\'aros, P., Rees, M. J., 1998. ApJ 503, 314

\bibitem[Ramirez-Ruiz et al.(2001a)]{enrico} Ramirez-Ruiz, E., 
  Dray, L. M., Madau, P., Tout, C. A., 2001a. MNRAS 327, 829 

\bibitem[Ramirez-Ruiz et al.(2001b)]{rmr01} Ramirez-Ruiz, E.,
Merloni A., Rees M. J., 2001b. MNRAS 324, 1147

\bibitem[Rees \& M\'{e}sz\'{a}ros(1998)]{rm98} Rees, M. J., 
M\'{e}sz\'{a}ros, P., 1998. ApJ 496, L1

\bibitem[Rees \& M\'esz\'aros(2000)]{rees} Rees, M. J., 
M\'esz\'aros, P., 2000. ApJ 545, L73

\bibitem[Reeves et al.(2002)]{james} Reeves, J. N., et al., 2002. 
Nature 416, 512

\bibitem[Santos-Lleo et al.(2001)]{santos} Santos-Lleo, M., 
Loiseau, N., Rodriguez, P., Altieri, B., Schartel, N., 2001. 
GCN Circ. 1192

\bibitem[Stanek et al.(1999)]{stanek} Stanek, K. Z., Garnavich, P.,
  Kaluzny, J., Pych, W., Thompson, I., 1999. ApJ 522, L39 

\bibitem[Uemura et al.(2003)]{japan} Uemura, M., et al., 2003. 
Nature 423, 843

\bibitem[van Paradijs et al.(2000)]{pkw} van Paradijs, J., 
Kouveliotou, C., Wijers, R. A. M. J., 2000. ARA\&A 38, 379

\bibitem[Vietri \& Stella(1999)]{vs} Vietri, M., Stella, L., 1999. 
ApJ 527, L43

\bibitem[Wang \& Loeb(2000)]{wl} Wang, X., Loeb, A., 2000. 
ApJ 535, 788

\bibitem[Watson et al.(2003)]{darach} Watson, D., Reeves, J. N.,
  Hjorth, J., Jakobsson, P., Pedersen, K., 2003. ApJ 595, L29

\bibitem[Wheeler et al.(2000)]{wheeler} Wheeler, J. C., Yi, I., 
Hoefflich, P., Wang, L., 2000. ApJ 537, 810

\bibitem[Zhang et al.(2003)]{zhang} Zhang,~W.,
  Woosley,~S.~E., MacFadyen,~A.~I., 2003. ApJ 586, 356

\end{thebibliography}
\end{document}